\documentclass[a4paper,11pt]{article}

\usepackage{booktabs}

\usepackage{lineno}
\usepackage[colorlinks=false]{hyperref}
\usepackage{bm}
\usepackage{graphicx}
\usepackage{amssymb}
\usepackage{amsmath}
\usepackage{tabularx}
\usepackage{natbib}
\usepackage{floatrow}
\usepackage{caption}
\usepackage{multirow}
\usepackage{amsmath}
\usepackage{amsfonts}
\usepackage{amssymb}
\usepackage{xcolor}
\usepackage{subfig}
\usepackage{graphicx,amsmath,bm}
\usepackage[binary-units=true]{siunitx}
\usepackage{array}
\usepackage{authblk}
\usepackage{rotating}
\usepackage{enumerate}
\usepackage{ragged2e}

\usepackage[left=15mm,right=15mm,top=1.5cm,bottom=1.5cm,includeheadfoot]{geometry}
\setcitestyle{square,numbers}
\begin{document}
	
	\title{Coordinate-Residual Physics-Driven Neural Network for Inverse Scattering Imaging}
	
	\author[1]{Yutong~Du}
	\author[1]{Zicheng~Liu}
	\author[1]{Bo~Qi}
	\author[1]{Yali~Zong}
	\author[1]{Peixian~Han}
	
	\affil[1]{\scriptsize Department of Electronic Engineering, School of Electronics and Information, Northwestern Polytechnical University, Xi'an 710029, China}
	\maketitle
	
\abstract
	{
		Electromagnetic inverse scattering is a nonlinear and ill-posed computational imaging problem, where accurate reconstruction is challenging due to measurement limitations, noise, and high computational costs, especially for 3-D imaging. Although physics-driven neural networks (PDNNs) reduce the dependence on labeled training data, existing accelerated PDNN frameworks often rely on preliminary reconstruction-based region selection, which may introduce instability when the selected region is inaccurate. In this paper, a coordinate-residual physics-driven neural network (CRPDNN) is proposed for 3-D electromagnetic inverse scattering. CRPDNN represents the unknown complex contrast distribution using normalized spatial coordinates and a residual convolutional network, whose parameters are optimized by enforcing consistency between the measured and model-predicted scattered fields. Unlike existing subregion-accelerated PDNN approaches, CRPDNN does not require a preliminary reconstruction, thereby avoiding dependence on its accuracy. For the reported noise-free 3-D synthetic cases, CRPDNN achieves an average relative error of 2.10\%, compared with 7.97\% for CSI and 3.99\% for $L_{2/3}$-FBE-WCIE, while providing approximately 5.5- and 12.1-fold speedups over the two baselines, respectively. Additional 2-D comparisons further demonstrate its stability and computational efficiency relative to existing PDNN frameworks. CRPDNN also maintains reliable reconstruction performance under noisy measurements, and the 3-D Fresnel experiments further indicate its potential for practical imaging applications. 
	}
	
	\section{Introduction}
	Electromagnetic inverse scattering~\cite{chen2018computationalEMIS} aims to reconstruct the dielectric properties of unknown objects from measured scattered fields. It has applications in nondestructive testing \cite{Amineh2024pipe}, computational microscopy \cite{Qin2026microscopy,liu2025computational}, biomedical imaging \cite{Guo2026BioMedical}, geophysical sensing \cite{Song2026geophysical}, and microwave imaging \cite{Du2026PDF}. However, the problem is nonlinear because the total field inside the object depends on the unknown material distribution. It is also ill-posed because the available scattered field measurements provide only limited and incomplete information about a large number of unknown contrast variables after discretization. These difficulties become more pronounced in 3-D volumetric imaging, where the number of unknowns increases substantially.
	
	Classical inversion methods can be roughly divided into linearized \cite{habashyi1993BA,devaney1982BP,devaney1981RA} and nonlinear iterative approaches \cite{sung1999BIM,peter1997CSI,chen2010SOM,chen_2010SOM,pan2011SOM}. Linearized methods, such as Born approximation (BA) \cite{habashyi1993BA,gao2006BA} and back-propagation (BP) \cite{devaney1982BP,tsili1998BP}, are computationally efficient and often used to provide initial estimates, but they are limited to weak-scattering scenarios. Nonlinear iterative methods, such as the distorted Born iterative method (DBIM) \cite{sung1999BIM}, contrast source inversion (CSI) \cite{peter1997CSI}, and the subspace-based optimization method (SOM) \cite{chen2010SOM,chen_2010SOM,pan2011SOM}, can better handle multiple scattering effects than linearized methods. However, their reconstruction performance can be sensitive to initialization, regularization settings, and measurement noise. Moreover, they often require considerable computation time, especially for large-scale 3-D inverse scattering problems.
	
	Deep learning \cite{wei2018BPS,Li2018DeepNIS,wei2019PhaNN,Liu2022PhaGuideNN,Liu2022SOMnet,Tao2023NNBIM,Zhang2023CSINet,Du2025QuaDNN} has been introduced into inverse scattering to learn a direct mapping from measured data or preliminary reconstructions to the target contrast. Once trained, these schemes can provide fast inference. However, their performance strongly depends on the training distribution. When the measurement configuration, noise level, or target class deviates from those covered by the training data, their generalization ability may degrade. Moreover, obtaining representative labeled datasets for 3-D electromagnetic inverse scattering is usually expensive because each training sample requires solving a 3-D forward scattering problem.
	
	Physics-driven neural networks (PDNNs) \cite{Du2025PDNN,Du2026IPDNN} have been developed to reduce the dependence on labeled training datasets. Instead of learning a general mapping from a large training set, PDNNs optimize the network parameters for each measurement case by minimizing a data-consistency loss. However, existing PDNN-based frameworks still have several limitations. First, existing subregion-accelerated PDNN frameworks usually require a preliminary reconstruction, such as BP, to guide the selection of a reduced reconstruction region. Although this strategy can reduce the computational burden, its performance becomes sensitive to the reliability of the preliminary estimate and the generalization ability of the subregion-selection model. Second, previous 2-D PDNN implementations were designed for planar imaging problems and contain fully connected components associated with the discretized reconstruction domain. A direct extension of such architectures to 3-D volumetric imaging is not straightforward, since the number of unknown voxels increases substantially and the corresponding memory and parameter costs may become prohibitive.
	
	To address these limitations, this work proposes a coordinate-residual physics-driven neural network (CRPDNN) for 3-D electromagnetic inverse scattering. CRPDNN represents the unknown complex contrast distribution using normalized spatial coordinates and a residual convolutional network, whose parameters are optimized by enforcing consistency between the predicted and measured scattered fields. Unlike existing accelerated PDNN frameworks, CRPDNN does not require a preliminary reconstruction, thereby avoiding the instability sensitivity to inaccurate preliminary estimates. The proposed solver is evaluated on 3-D synthetic and Fresnel experimental data, including complex-valued targets and noise-contaminated measurements. Additional 2-D comparisons with existing PDNN frameworks is also conducted to assess its performance under the original 2-D PDNN setting. The source code is publicly available at https://github.com/Physics-driven-methods/Coordinate-Residual-Physics-Driven-Neural-Network-for-Inverse-Scattering-Imaging.
	
	\begin{figure}
		\centering
		\includegraphics[width = .66\linewidth]{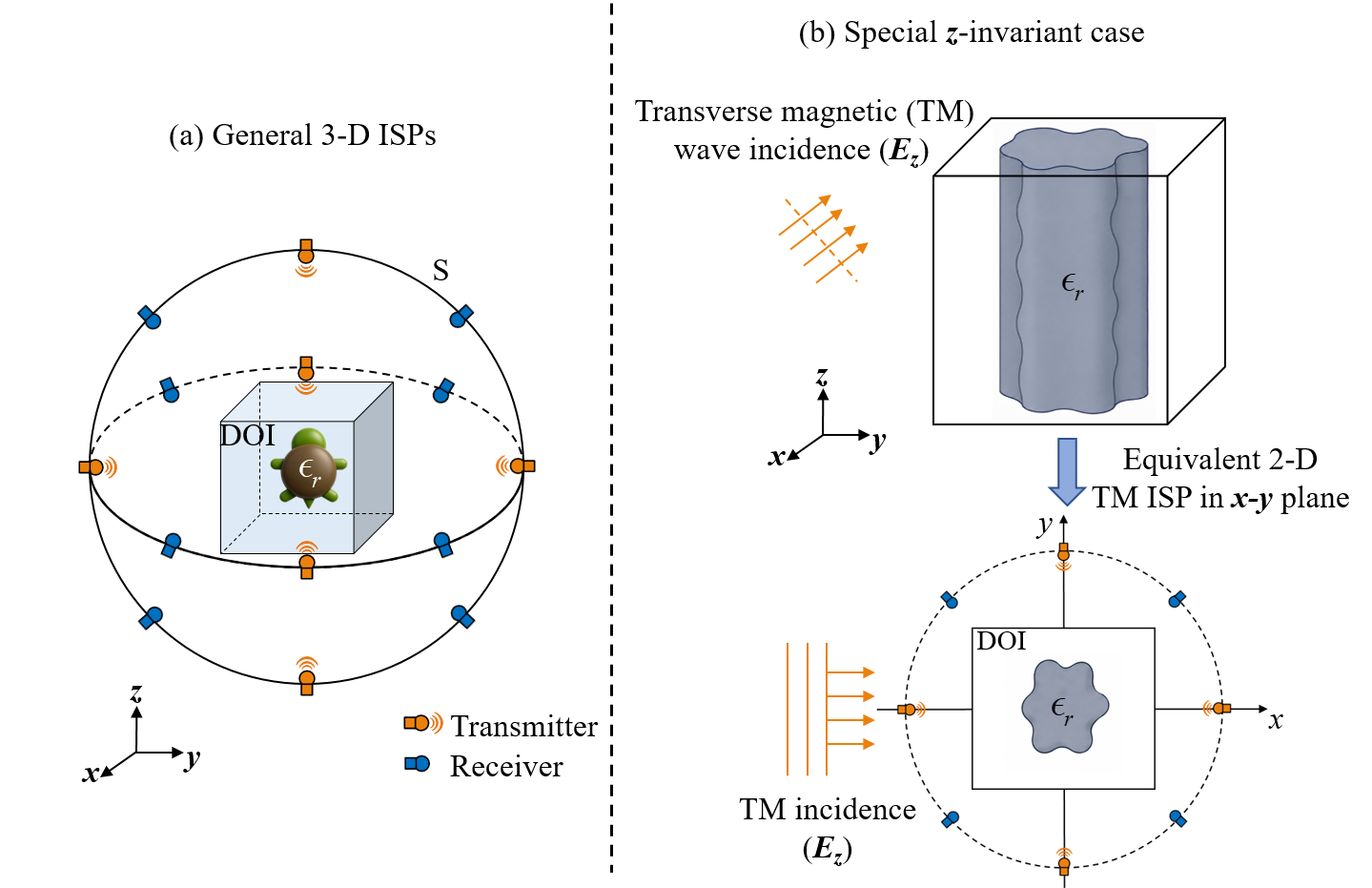}
		\caption{Schematic illustration of the considered inverse scattering problems (ISPs). (a) General 3-D ISP. (b) Special $z$-invariant TM case, where a $z$-invariant object under transverse magnetic (TM) incidence $(E_z)$ can be reduced to an equivalent 2-D TM inverse scattering problem in the $x$--$y$ plane.}
		\label{fig:ISPs}
	\end{figure}
	
	\section{Formulation of Electromagnetic Inverse Scattering Problems}
	\label{sec:formulateISP}
	
	In the considered electromagnetic ISPs, the domain of interest (DOI) is embedded in free space and sequentially illuminated by $N_i$ incident waves. The scattered fields generated by the unknown object are collected by $N_s$ receivers distributed on an observation surface $S$ surrounding the DOI, as illustrated in Figure~\ref{fig:ISPs} (a). The objective is to reconstruct the relative permittivity distribution of the targets from the measured scattered fields. 
	
	For the 3-D vector case, the interaction between the incident wave and the targets is described by the vector volume integral equation (VIE) \cite{Schaubert1984VIE,Roger1961THEMF}. The total electric field inside the DOI satisfies
	\begin{equation}
		\mathbf E^{\mathrm{tot}}(\mathbf r) = \mathbf E^{\mathrm{inc}}(\mathbf r) 
		+ k_0^2 \int_{\mathrm{DOI}} {\mathbf G}(\mathbf r,\mathbf r') 
		\mathbf J(\mathbf r') d\mathbf r',
		\quad \mathbf r\in \mathrm{DOI},
		\label{eq:state_equation_3-D}
	\end{equation}
	where $\mathbf E^{\mathrm{inc}}$ and $\mathbf E^{\mathrm{tot}}$ denote the incident and total electric fields, respectively. Here, $k_0$ is the background wavenumber, and ${\mathbf G}(\mathbf r,\mathbf r')$ denotes the dyadic Green's function \cite{tai1994dyadic}. The contrast source is defined as
	\begin{equation}
		\mathbf J(\mathbf r) = \chi(\mathbf r)\mathbf E^{\mathrm{tot}}(\mathbf r),
		\label{eq:contrast_source}
	\end{equation}
	where $\chi(\mathbf r)=\epsilon_r(\mathbf r)-1$ is the contrast function, with $\epsilon_r(\mathbf r)$ being the relative permittivity. The scattered field on the observation surface is radiated by the induced contrast source and is given by
	\begin{equation}
		\mathbf E^{\mathrm{sca}}(\mathbf r) = 
		k_0^2\int_{\mathrm{DOI}}{\mathbf G}(\mathbf r,\mathbf r')
		\mathbf J(\mathbf r')d\mathbf r',
		\quad \mathbf r\in S.
		\label{eq:data_equation_3-D_vector}
	\end{equation}
	
	For numerical implementation, the DOI is discretized into voxels for 3-D problems. The forward problem is solved by an FFT-accelerated BiCGSTAB scheme, where the dyadic Green's-function convolution and voxel self-interaction term are treated following the standard formulation \cite{Zhong2019FFT}.
	
	Given the measured scattered field data, the ISPs aim to reconstruct the unknown contrast distribution $\chi$ within the DOI. Due to the inherent nonlinearity and ill-posedness of the ISPs, suitable regularization or prior constraints are required to obtain stable and physically meaningful reconstructions\cite{chen2018computationalEMIS}.
	
	\begin{figure*}
		\centering
		\includegraphics[width = \linewidth]{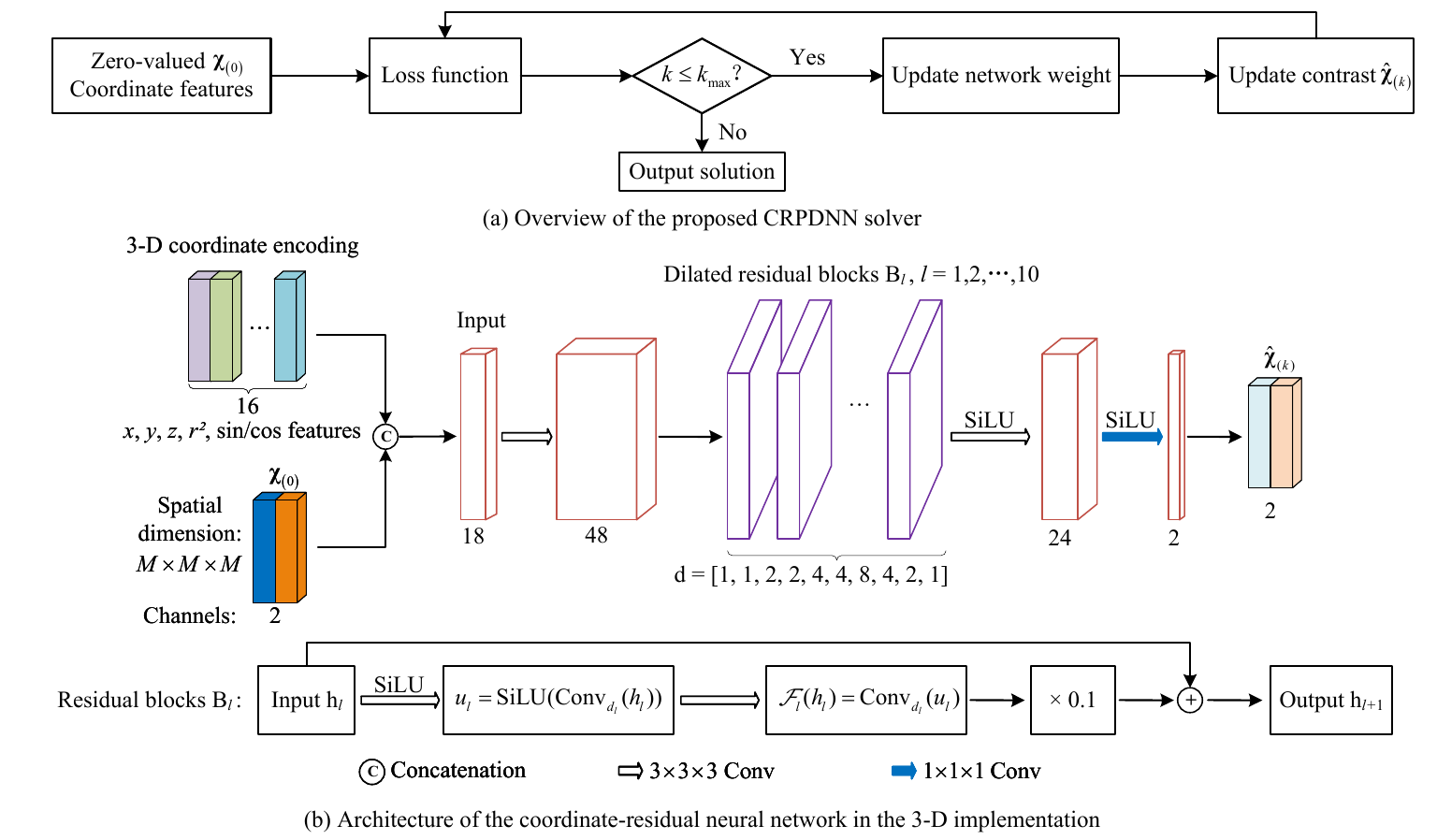}
		\caption{
			Sketch of the proposed coordinate-residual physics-driven neural network (CRPDNN) solver. 
			(a) Overall optimization framework of CRPDNN. 
			(b) Architecture of the coordinate-residual neural network in the 3-D implementation.}
		\label{fig:Overview}
	\end{figure*}
	
	\section{Coordinate-Residual Physics-Driven Neural Network Solver}
	\label{sec:CRPDNN}
	This work develops a coordinate-residual physics-driven neural network solver (CRPDNN) for electromagnetic inverse scattering. The overall framework of the proposed solver is shown in Figure~\ref{fig:Overview}(a). The network input consists of two zero-valued contrast channels and the coordinate features. Specifically, the real and imaginary channels of $\boldsymbol\chi_{(0)}$ are both initialized as zero and are not obtained from any fast preliminary reconstruction. The coordinate-residual network then generates the contrast estimate $\hat{\boldsymbol\chi}_{(k)}$, which is used to evaluate the loss function. Guided by the loss function, the network weights are iteratively updated until the maximum number of iterations is reached, and the final contrast estimate is taken as the reconstructed solution.
	
	\subsection{Coordinate-Residual Neural Network  Architecture}
	\label{subsec:CNNarc}
	The 3-D implementation of the proposed coordinate-residual neural network is shown in Figure~\ref{fig:Overview} (b). The network is designed to represent the unknown complex contrast distribution. To provide explicit spatial information, a 3-D coordinate encoding is concatenated with the initial zero-value contrast. The coordinates of the reconstruction domain are first normalized to $[-1,1]$ along the three spatial directions. For each voxel position $\mathbf r=(x,y,z)$, the coordinate encoding is defined as
	\begin{equation}
		\gamma(\mathbf r)=\Big\{x,y,z,r^2,\sin(n\pi \xi),\cos(n\pi \xi)\Big\}_{\xi\in\{x,y,z\},\,n=1,2}
	\end{equation}
	where $r^2=x^2+y^2+z^2$. Therefore, the coordinate encoding $\gamma(\mathbf r)$ consists of 16 feature channels. The network input is obtained by concatenating the 2 channel initial contrast and the 16 channel coordinate encoding along the channel dimension. The 18 channel input is first projected to a 48 channel latent feature volume by a $3\times3\times3$ convolution with padding 1. The latent feature tensor is then processed by ten dilated residual blocks $\mathbf B_l$. The dilation rates of these blocks are set as $[1,1,2,2,4,4,8,4,2,1]$. For the $l$-th block, the residual mapping consists of two $3\times3\times3$ convolutional layers with the same dilation rate $d_l$, and a SiLU activation applied after the first convolution. The padding is set to $d_l$ so that the spatial size is preserved. The block output is given by $\mathbf h_l+0.1\mathcal F_l(\mathbf h_l)$, where $\mathcal F_l(\cdot)$ denotes the two-layer dilated convolutional residual function. The residual scaling factor is set to $0.1$ for stable feature updating. The use of different dilation rates enlarges the receptive field while preserving the spatial resolution and convolution kernel size. After the residual blocks, the feature tensor is passed through a SiLU activation and a $3\times3\times3$ convolution to reduce the feature channels from 48 to 24, followed by another SiLU activation and a $1\times1\times1$ convolution to generate the two-channel complex contrast estimate $\hat{\boldsymbol{\chi}}_{k}$.
	
	The 2-D network architecture follows the same coordinate-residual design and can be regarded as a dimensionally reduced version of the 3-D network, where the $z$ coordinate is removed and all $3\times3\times3$ convolutions are replaced by $3\times3$ convolutions. For a 2-D position $\mathbf r=(x,y)$, the coordinate encoding is defined as
	\begin{equation}
		\gamma_{\mathrm{2D}}(\mathbf r)=
		\Big\{x,y,r^2,\sin(n\pi \xi),\cos(n\pi \xi)\Big\}_{\xi\in\{x,y\},\,n=1,2}
	\end{equation}
	where $r^2=x^2+y^2$. Thus, the 2-D coordinate encoding contains 11 feature channels. Except for this dimensional change, the residual representation and optimization strategy are kept consistent with the 3-D implementation.
	
	\subsection{Loss Functions}
	\label{subsec:lossfun}
	The total loss function is composed of three terms
	\begin{equation}
		\text{Loss} = L_{\text{Data}}+\lambda_{\mathrm{Bound}} L_{\text{Bound}}+\lambda_{\mathrm{TV}} L_{\text{TV}}
		\label{LossFunc}
	\end{equation}
	where $\lambda_{\mathrm{Bound}}$ and $\lambda_{\mathrm{TV}}$ are the weighting coefficients determined by the scale-balanced strategy described in Section~\ref{subsec:weight}.
	
	The data consistency term is defined as 
	\begin{equation}
		L_{\text{Data}}=\frac{||{\mathbf{E}}^\text{sca}_\text{mea}-\hat{\mathbf{E}}^\text{sca}({\hat{\boldsymbol{\chi}}})||_2}{||{\mathbf{E}}^\text{sca}_\text{mea}||_2},
		\label{eq:LData}
	\end{equation}
	where $\mathbf{E}^{\mathrm{sca}}_{\mathrm{mea}}$ and 
	$\hat{\mathbf{E}}^{\mathrm{sca}}(\hat{\boldsymbol{\chi}})$ denote the measured scattered fields and the scattered fields computed from the predicted contrast $\hat{\boldsymbol{\chi}}$, respectively.
	
	The lower-bound constraint term is written as
	\begin{equation}
		L_{\text{Bound}}=\frac{1}{N}\sum_{n=1}^{N}\text{ReLU}(-\mathbf{Re(\hat{\boldsymbol{\chi}})}),
		\label{eq:LreEp}
	\end{equation}
	where $N$ is the number of voxels in the DOI. This term suppresses nonphysical negative values in the real part of the reconstructed contrast.
	
	The TV regularization term is imposed on the magnitude of the reconstructed complex contrast
	\begin{equation}
		L_{\mathrm{TV}}=\frac{1}{N}\sum_{i,j,k}\sqrt{(\nabla_x|\hat{\boldsymbol{\chi}}_{i,j,k}|)^2+(\nabla_y |\hat{\boldsymbol{\chi}}_{i,j,k}|)^2+(\nabla_z |\hat{\boldsymbol{\chi}}_{i,j,k}|)^2}.
		\label{eq:LTV3D}
	\end{equation}
	Here, $\nabla_{\alpha}|\hat{\boldsymbol{\chi}}_{i,j,k}|$ denotes the forward finite difference of $|\hat{\boldsymbol{\chi}}|$ along the $\alpha$-direction, with $\alpha\in\{x,y,z\}$. This term suppresses local oscillations and promotes segmentation uniformity, while still allowing sharp material boundaries. For the 2-D implementation, the TV term is reduced accordingly and is evaluated only along the $x$- and $y$-directions.
	
	\subsection{Scale-Balanced Adaptive Weighting}
	\label{subsec:weight}
	The $L_{\text{Bound}}$ and $L_{\text{TV}}$ have different numerical scales from the $L_{\text{Data}}$. Directly assigning fixed empirical weights may therefore lead to sample-dependent behavior, especially when the noise level of the measured data is unknown. Therefore, a scale-balanced adaptive weighting strategy is used.
	
	The weights are computed at the $k$-th iteration according to the loss scales
	\begin{equation}
		\lambda_i^{(k)}=\rho_i \frac{L^{(k)}_{\mathrm{Data}}}{L_i^{(k)}+\epsilon}, \quad i\in\{\mathrm{Bound},\mathrm{TV}\},
		\label{eq:adaptive_weight}
	\end{equation}
	where $\epsilon=1\times10^{-8}$ ensures numerical stability. In this work, $\rho_{\mathrm{Bound}}=0.05$ and $\rho_{\mathrm{TV}}=0.02$ are used for all reconstructions. In implementation, when $L^{\mathrm{Bound}}$ is below $\epsilon$, $\lambda_{\mathrm{Bound}}^{(k)}$ is set to zero to avoid unnecessary amplification of an already inactive bound penalty.
	
	\subsection{Training Settings}
	\label{subsec:trainSet}
	All methods were implemented and evaluated on a workstation equipped with 128 GB RAM, a 3.2 GHz Intel i9 CPU and an NVIDIA GeForce RTX 4090 GPU. The CRPDNN parameters were optimized using Adam algorithm with an initial learning rate of 1$\times$10$^{-3}$, which is halved every 150 iterations. Training is performed for a maximum of 300 iterations.
	
	\begin{figure*}
		\centering
		\includegraphics[width = \linewidth]{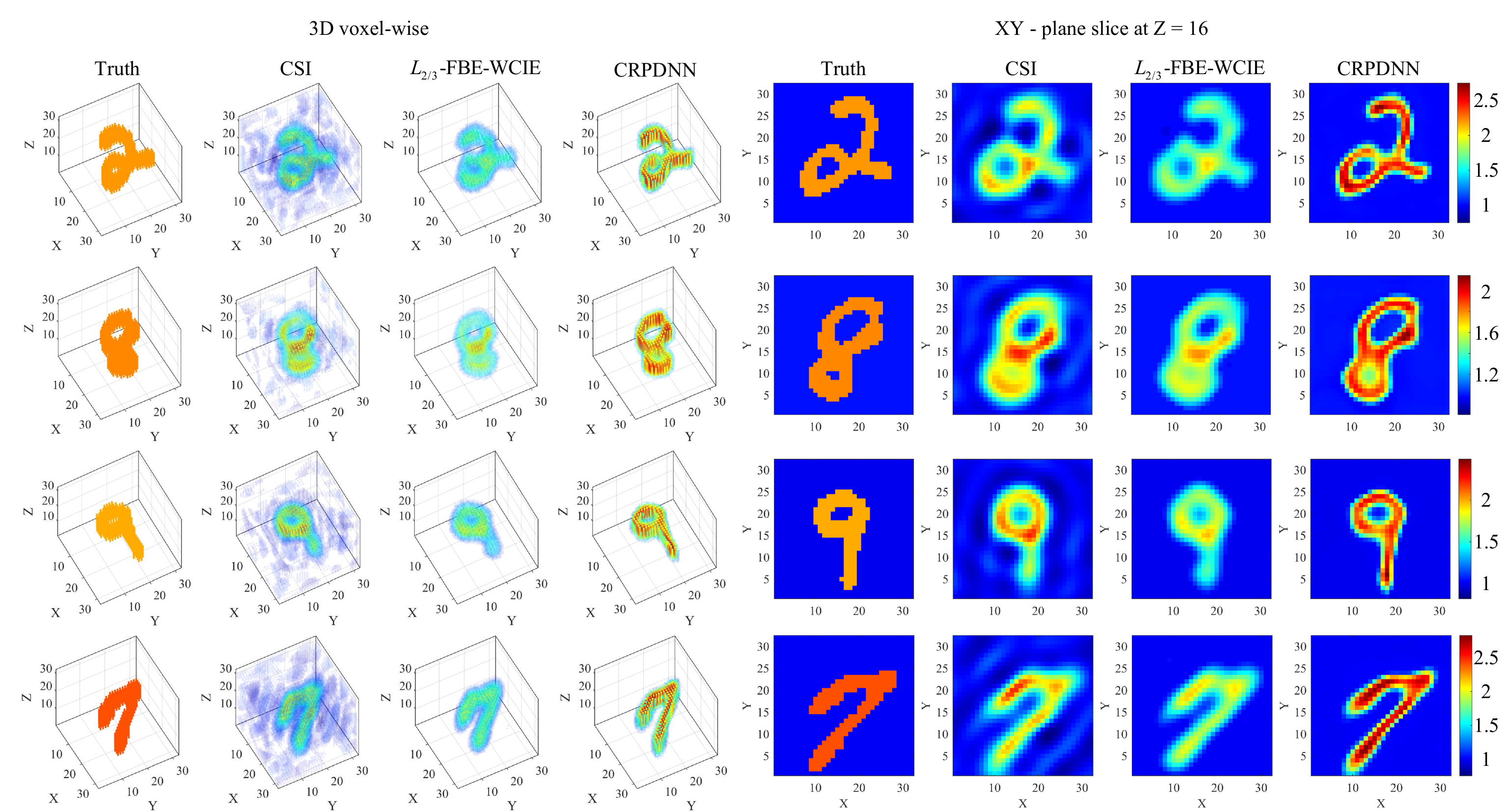}
		\caption{Reconstruction results for four 3-D digit-like samples obtained using the proposed CRPDNN and the baseline methods CSI and $L_{2/3}$-FBE-WCIE.}
		\label{fig:MNIST3D}
	\end{figure*}
	
	\section{Numerical Results}
	\label{sec:numAna}
	
	In this section, the proposed CRPDNN solver is mainly evaluated on 3-D inverse scattering problems, including digit-like samples, complex-valued profiles and noisy measurements. CSI and $L_{2/3}$-FBE-WCIE \cite{Li2026FBEL23} are used as baseline methods for quantitative and visual comparison. In addition, a 2-D comparison with existing PDNN frameworks is provided to demonstrate the advantage of CRPDNN in avoiding unreliable subregion selection.
	
	For the 3-D synthetic experiments, the operating frequency is set to 4 GHz. The DOI is a $0.15\,\mathrm{m}\times0.15\,\mathrm{m}\times0.15\,\mathrm{m}$ cubic domain discretized into a $32\times32\times32$ voxel grid. The object is illuminated by 16 incident waves from different directions, and the scattered fields are collected by 32 receivers distributed on a spherical observation surface with radius $20\lambda$, where $\lambda$ is the free-space wavelength. Both the transmit and receive polarizations are set to the $\phi$ direction, corresponding to a co-polarized $\phi\phi$ configuration. In the noise tests, additive white Gaussian noise (AWGN) is added to the scattered data.
	
	The 3-D results are visualized by voxel-based scatter plots with contrast-dependent opacity. For $\operatorname{Re}(\epsilon_r)$, the opacity values are assigned according to $|\operatorname{Re}(\epsilon_r)-1|$ using thresholds of 0.05, 0.20, and 0.60, with opacity values of 0.08, 0.25, and 1.00, respectively. For $\operatorname{Im}(\epsilon_r)$, the thresholds are set to 0.01, 0.05, and 0.15, with opacity values of 0.05, 0.20, and 0.80, respectively. The 2-D slice views of the 3-D results are directly extracted from the reconstructed volumes without post-processing.
	
	\subsection{3-D digit-like profiles}
	\label{subsec:MNIST3D}
	Figure~\ref{fig:MNIST3D} shows the reconstruction results of four 3-D digit-like samples. Both the 3-D voxel-wise visualization and the central slice at $Z=16$ are presented for comparison. CSI can approximately locate the targets, but its reconstructions contain noticeable background artifacts and significantly underestimate the target contrast. The $L_{2/3}$-FBE-WCIE method suppresses part of the background artifacts and improves the compactness of the reconstructed targets. However, the reconstructed profiles remain overly smooth, with attenuated fine structures.
	
	In comparison, CRPDNN provides more accurate 3-D reconstructions. The reconstructed digit shapes are clearer in both the volumetric and the 2-D slice views. In particular, thin strokes and curved boundaries are better preserved, while background artifacts are effectively suppressed.
	
	The quantitative results in Table~\ref{tab:MNIST3D} are consistent with the visual comparison. CRPDNN achieves the lowest relative error in all four cases. The average relative errors of CSI, $L_{2/3}$-FBE-WCIE, and CRPDNN are 7.29\%, 3.36\%, and 1.98\%, respectively. In terms of runtime, CSI and $L_{2/3}$-FBE-WCIE require about 711 s and 1643 s on average, respectively, while CRPDNN only requires about 133 s. These results show that CRPDNN provides higher reconstruction accuracy and substantially lower runtime for the tested 3-D digit-like samples.
	
	\begin{table}[!ht]
		\centering
		\caption{Quantitative comparison of relative error and runtime for the 3-D digit-like samples}
		\label{tab:MNIST3D}
		\resizebox{0.66\columnwidth}{!}{
			\begin{tabular}{l|c c c| c c c}
				\toprule
				\textbf{} & \multicolumn{3}{c|}{\textbf{Relative error}} & \multicolumn{3}{c}{\textbf{Runtime}} \\ 
				\midrule
				\textbf{Methods} & \textbf{CSI} & \textbf{$L_{2/3}$-FBE-WCIE} & \textbf{CRPDNN} & \textbf{CSI} & \textbf{$L_{2/3}$-FBE-WCIE} & \textbf{CRPDNN} \\ 
				\midrule
				Case 1 &8.60\%  &4.23\%   &2.36\% &708~s  &1596~s  &160~s \\ 
				Case 2 &5.68\%  &2.65\%   &1.87\% &707~s  &1806~s  &108~s \\ 
				Case 3 &5.67\%  &2.42\%   &1.35\% &717~s  &1585~s  &116~s \\ 
				Case 4 &9.22\%  &4.12\%   &2.32\% &713~s  &1585~s  &146~s \\ 
				\bottomrule
			\end{tabular}
		}
	\end{table}
	
	\subsection{3D complex profiles}
	\label{subsec:Complex3D}
	
	\begin{figure}[!t]
		\centering
		\includegraphics[width = \linewidth]{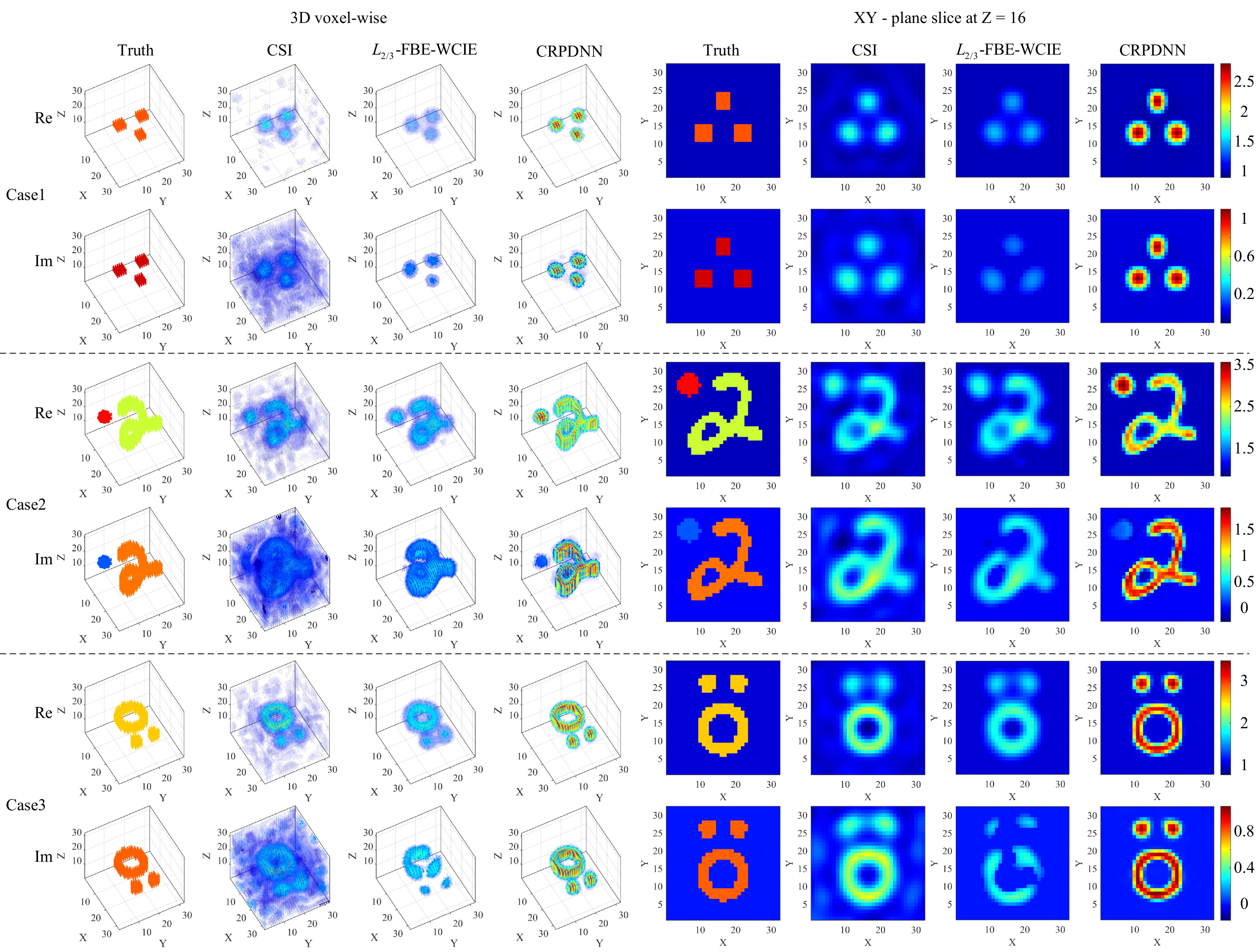}
		\caption{Reconstruction results for three 3-D complex samples obtained using the proposed CRPDNN and the baseline methods CSI and $L_{2/3}$-FBE-WCIE.}
		\label{fig:Complex3D}
	\end{figure}
	
	\begin{figure}[!t]
		\centering
		\includegraphics[width = \linewidth]{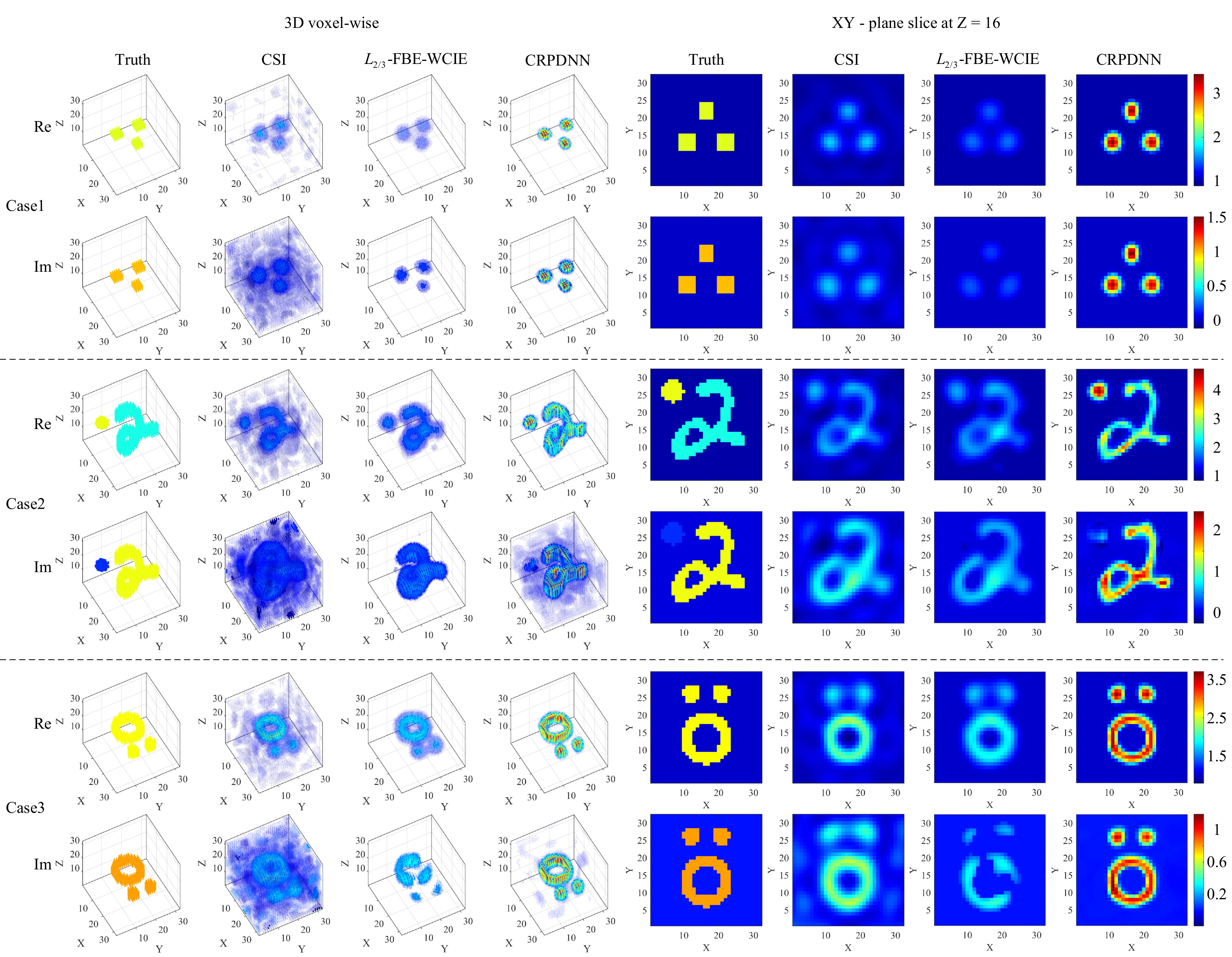}
		\caption{Reconstruction results for three 3-D complex samples obtained using the proposed CRPDNN and the baseline methods CSI and $L_{2/3}$-FBE-WCIE under AWGN with SNR = 20 dB.}
		\label{fig:Noise3DSNR20dB}
	\end{figure}
	
	\begin{figure}[!t]
		\centering
		\includegraphics[width = \linewidth]{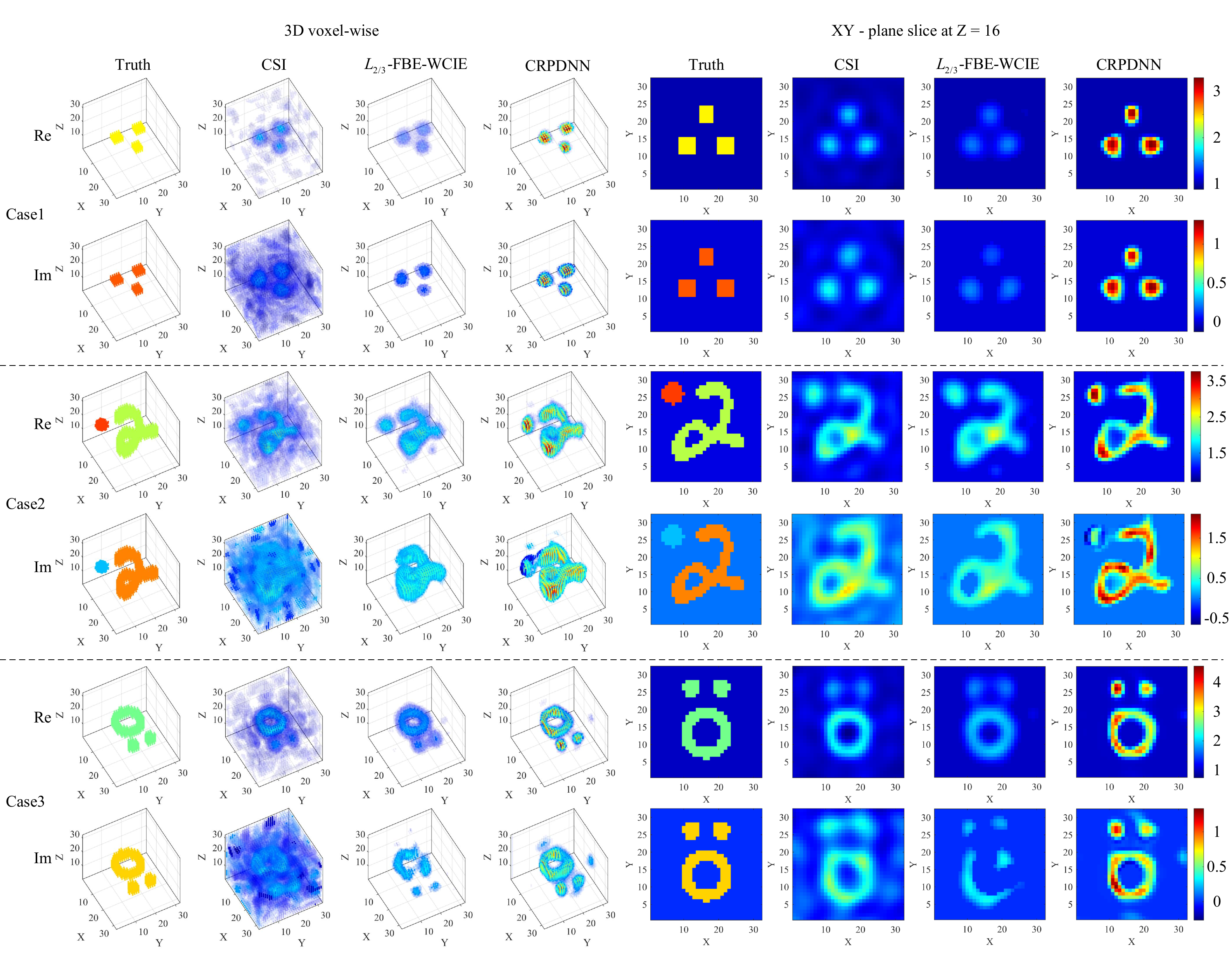}
		\caption{Reconstruction results for three 3-D complex samples obtained using the proposed CRPDNN and the baseline methods CSI and $L_{2/3}$-FBE-WCIE under AWGN with SNR = 10 dB.}
		\label{fig:Noise3DSNR10dB}
	\end{figure}
	
	Figure~\ref{fig:Complex3D} shows the reconstruction results of three 3-D complex-valued profiles. Compared with the 3-D digit-like samples, these cases are more challenging because the real and imaginary parts of the relative permittivity need to be reconstructed simultaneously.
	
	For Case 1, all methods can identify the three separated parts. However, CSI produces noticeable background artifacts, while $L_{2/3}$-FBE-WCIE gives more compact but slightly weakened targets. CRPDNN better preserves the location, size, and contrast of the three parts in both the real and imaginary parts.
	
	For Case 2, CSI recovers the rough target profile but suffers from blurring and background artifacts. The $L_{2/3}$-FBE-WCIE method reduces background artifacts, but its sparsity promoting constraint also suppresses the weak spherical scatterer in the imaginary part, indicating a risk of removing physically meaningful weak scatterers. In contrast, CRPDNN better preserves both the main digit-like structure and the weak scatterer, resulting in a more complete complex-valued reconstruction.
	
	For Case 3, CSI roughly locates the targets but produces blurred boundaries and noticeable background artifacts. The $L_{2/3}$-FBE-WCIE method reduces background artifacts, but the reconstructed result exhibits partial loss of the ring structure, which may be associated with the strong sparsity-promoting regularization. In contrast, CRPDNN better preserves the ring topology and the two small scatterers, with clearer boundaries.
	
	\begin{table}[!ht]
		\centering
		\caption{Quantitative comparison of relative error and runtime for the 3-D complex samples}
		\label{tab:Complex3D}
		\resizebox{0.66\columnwidth}{!}{
			\begin{tabular}{l|c c c| c c c}
				\toprule
				\textbf{} & \multicolumn{3}{c|}{\textbf{Relative error}} & \multicolumn{3}{c}{\textbf{Runtime}} \\ 
				\midrule
				\textbf{Methods} & \textbf{CSI} & \textbf{$L_{2/3}$-FBE-WCIE} & \textbf{CRPDNN} & \textbf{CSI} & \textbf{$L_{2/3}$-FBE-WCIE} & \textbf{CRPDNN} \\ 
				\midrule
				Case 1 &4.42\%  &1.60\%   &1.09\% &760~s  &1574~s  &72~s \\ 
				Case 2 &13.42\%  &8.23\%   &3.43\% &715~s  &1575~s  &185~s \\ 
				Case 3 &8.78\%  &4.71\%   &2.29\% &780~s  &1575~s  &144~s \\ 
				\bottomrule
			\end{tabular}
		}
	\end{table}
	
	The quantitative results in Table~\ref{tab:Complex3D} are consistent with the visual comparison. CRPDNN achieves the lowest relative error in all three cases. The average relative errors of CSI, $L_{2/3}$-FBE-WCIE, and CRPDNN are 8.87\%, 4.85\%, and 2.27\%, respectively. CRPDNN shows the largest improvement in Case 2. In terms of runtime, CRPDNN requires about 134 s on average, which is much shorter than CSI and $L_{2/3}$-FBE-WCIE, whose average runtimes are about 752 s and 1575 s, respectively. These results indicate that CRPDNN provides a favorable balance between complex-valued 3-D reconstruction accuracy and computational efficiency.
	
	\subsection{3D noise robustness}
	\label{subsec:noise3D}
	To evaluate the robustness of different methods to measurement noise, AWGN is added to the scattered field data with SNRs of 20 dB and 10 dB. The reconstruction results are shown in Figure~\ref{fig:Noise3DSNR20dB} and Figure~\ref{fig:Noise3DSNR10dB}. The quantitative comparisons are summarized in Table~\ref{tab:noise3D20dB} and Table~\ref{tab:noise3D10dB}.
	
	Under 20 dB noise, CSI can still recover the approximate object locations, but the reconstructed 3-D profiles contain noticeable background artifacts and blurred boundaries. The $L_{2/3}$-FBE-WCIE method suppresses background responses more effectively, but weak structures may also be over-regularized. In comparison, CRPDNN provides more compact object regions and clearer boundaries. For Case 2 and Case 3, CRPDNN better preserves the weak scatterer and ring-like structures, while the baseline methods show stronger background artifacts or partial loss of target components.
	
	When the SNR decreases to 10 dB, the reconstruction becomes more difficult for all methods. CSI exhibits stronger background fluctuations, and the reconstructed profiles become more blurred. The $L_{2/3}$-FBE-WCIE method remains effective in reducing background artifacts, but its sparsity-promoting constraint can further suppress weak but meaningful structures. CRPDNN also shows some degradation, but it still maintains clearer object shapes and better structural completeness than the two baseline methods.
	
	The quantitative results are consistent with the visual observations. At 20 dB, the average relative errors of CSI, $L_{2/3}$-FBE-WCIE, and CRPDNN are 9.10\%, 4.85\%, and 2.99\%, respectively. At 10 dB, the corresponding average errors are 10.49\%, 4.93\%, and 3.51\%. Although the error of CRPDNN increases as the noise level becomes stronger, it achieves the lowest relative error in all cases. In terms of runtime, CRPDNN requires about 158 s and 167 s on average under 20 dB and 10 dB noise, respectively, which is much shorter than the two baseline methods. These results indicate that CRPDNN maintains a favorable balance between noise robustness, reconstruction accuracy, and computational efficiency for 3-D complex-valued imaging.
	
	\begin{table}[!ht]
		\centering
		\caption{Quantitative comparison of relative error and runtime for the 3-D complex samples under AWGN with SNR = 20 dB}
		\label{tab:noise3D20dB}
		\resizebox{0.66\columnwidth}{!}{
			\begin{tabular}{l|c c c| c c c}
				\toprule
				\textbf{} & \multicolumn{3}{c|}{\textbf{Relative error}} & \multicolumn{3}{c}{\textbf{Runtime}} \\ 
				\midrule
				\textbf{Methods} & \textbf{CSI} & \textbf{$L_{2/3}$-FBE-WCIE} & \textbf{CRPDNN} & \textbf{CSI} & \textbf{$L_{2/3}$-FBE-WCIE} & \textbf{CRPDNN} \\ 
				\midrule
				Case 1 &4.50\%  &1.59\%   &0.86\% &921~s  &1576~s  &105~s \\ 
				Case 2 &13.73\%  &8.26\%   &5.37\% &870~s  &1573~s  &222~s \\ 
				Case 3 &9.07\%  &4.70\%   &2.73\% &810~s  &1572~s  &148~s \\ 
				\bottomrule
			\end{tabular}
		}
	\end{table}
	
	\begin{table}[!ht]
		\centering
		\caption{Quantitative comparison of relative error and runtime for the 3-D complex samples under AWGN with SNR = 10 dB}
		\label{tab:noise3D10dB}
		\resizebox{0.66\columnwidth}{!}{
			\begin{tabular}{l|c c c| c c c}
				\toprule
				\textbf{} & \multicolumn{3}{c|}{\textbf{Relative error}} & \multicolumn{3}{c}{\textbf{Runtime}} \\ 
				\midrule
				\textbf{Methods} & \textbf{CSI} & \textbf{$L_{2/3}$-FBE-WCIE} & \textbf{CRPDNN} & \textbf{CSI} & \textbf{$L_{2/3}$-FBE-WCIE} & \textbf{CRPDNN} \\ 
				\midrule
				Case 1 &5.13\%  &1.60\%   &0.93\% &897~s  &1576~s  &121~s \\ 
				Case 2 &15.55\%  &8.38\%   &6.24\% &942~s  &1584~s  &208~s \\ 
				Case 3 &10.78\%  &4.82\%   &3.35\% &855~s  &1571~s  &173~s \\ 
				\bottomrule
			\end{tabular}
		}
	\end{table}
	
	\subsection{Comparison With PDNN Frameworks}
	\label{subsec:comparePDNN}
	\begin{figure}[!t]
		\centering
		\includegraphics[width = .8\linewidth]{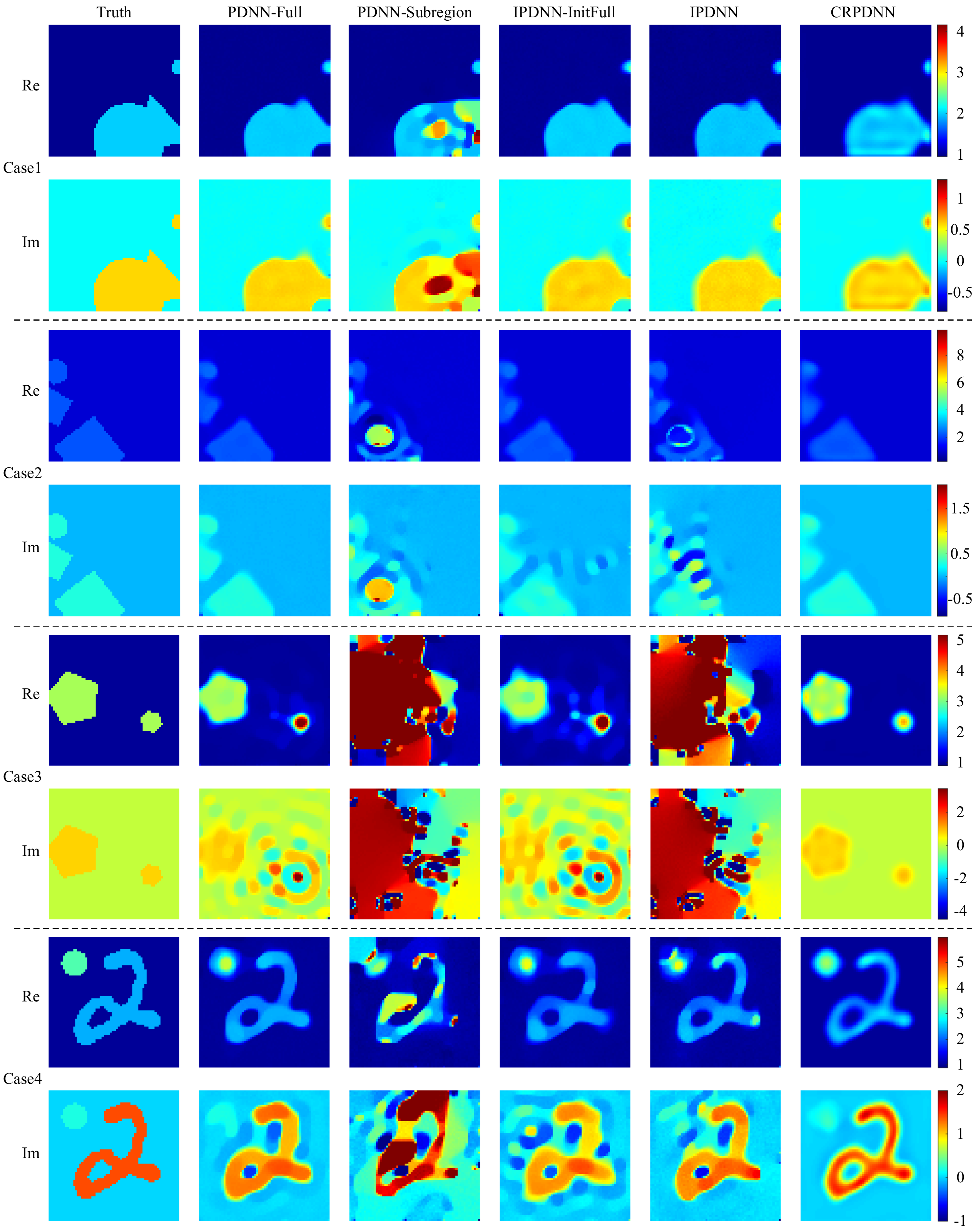}
		\caption{Reconstruction comparison between the proposed CRPDNN and existing PDNN-based frameworks.}
		\label{fig:comparePDNN}
	\end{figure}
	
	To further evaluate the stability of the proposed CRPDNN, it is compared with several existing PDNN-based frameworks \cite{Du2025PDNN,Du2026IPDNN}, including PDNN-Full, PDNN-Subregion, IPDNN-InitFull, and IPDNN. Since these baseline frameworks were originally proposed and mainly evaluated for 2-D transverse magnetic (TM) inverse scattering problems, the comparison is conducted using the same 2-D configuration as the baseline solvers for fair evaluation, as illustrated in Figure~\ref{fig:ISPs}(b).
	
	PDNN-Full and PDNN-Subregion are implemented based on the original PDNN framework with 3000 iterations, following the settings reported in the original work. The difference is that PDNN-Full performs the reconstruction over the whole DOI, whereas PDNN-Subregion first inputs the BP result into a U-Net and then selects a subregion by threshold segmentation for subsequent reconstruction. IPDNN-InitFull and IPDNN are both based on the IPDNN framework. In IPDNN-InitFull, the initial reconstruction region is set as the whole DOI, and the subregion is dynamically identified during the iterative process. In IPDNN, the initial subregion is the same as that in PDNN-Subregion, and the subsequent dynamic subregion identification follows the same strategy as IPDNN-InitFull. Unlike these PDNN frameworks, CRPDNN does not rely on preliminary inversion.
	
	Figure~\ref{fig:comparePDNN} and Table~\ref{tab:comparePDNN} present the qualitative and quantitative comparisons between the proposed CRPDNN and existing PDNN-based frameworks. For Case 1, except for PDNN-Subregion, the compared solvers obtain reasonable reconstruction results. The poor performance of PDNN-Subregion is mainly caused by the limited out-of-distribution generalization of the U-Net-based subregion selector. When the selected region is inaccurate, the following PDNN optimization is restricted to an incorrect domain, leading to artifacts and increased error. Although IPDNN starts from the same initial subregion, its dynamic subregion identification can recover part of the missed regions during the iterative process, and therefore provides a better reconstruction than PDNN-Subregion.
	
	The instability becomes more significant in the more challenging cases. In Case 3, PDNN-Subregion and IPDNN fail to produce reliable reconstructions, with relative errors of 432.3\% and 375.76\%, respectively. This result shows that subregion-based acceleration may become unreliable when the initial subregion is seriously biased. PDNN-Full avoids this issue by optimizing over the whole DOI, but it requires much longer runtime. IPDNN-InitFull also starts from the full domain and improves the robustness compared with subregion-initialized methods, but it still depends on the stability of the subsequent dynamic region identification.
	
	In contrast, CRPDNN does not require a fast-algorithm initial solution or any explicit subregion selection. The reconstruction is performed directly through the proposed physics-driven network representation. In addition, the loss weights are adaptively adjusted during optimization, reducing the sensitivity to manually selected hyperparameters. It is worth noting that, even without subregion-based acceleration, CRPDNN requires only 8~s, which is much shorter than the compared PDNN baselines.
	
	\begin{table*}[!ht]
		\centering
		\caption{Quantitative comparison of relative error and runtime between the proposed CRPDNN and existing PDNN-based frameworks}
		\label{tab:comparePDNN}
		\resizebox{1.0\columnwidth}{!}{
			\begin{tabular}{l|c c c c c| c c c c c}
				\toprule
				\textbf{} & \multicolumn{5}{c|}{\textbf{Relative error}} & \multicolumn{5}{c}{\textbf{Runtime}} \\ 
				\midrule
				\textbf{Methods} & \textbf{PDNN-Full} & \textbf{PDNN-Subregion} & \textbf{IPDNN-InitFull} & \textbf{IPDNN} & \textbf{CRPDNN} & \textbf{PDNN-Full} & \textbf{PDNN-Subregion} & \textbf{IPDNN-InitFull} & \textbf{IPDNN} & \textbf{CRPDNN} \\ 
				\midrule
				Case 1 &2.23\%  &14.16\%  &3.23\%  &3.82\%  &4.16\% &451~s  &70~s  &140~s  &44~s  &8~s \\ 
				Case 2 &3.84\%  &21.73\%  &5.95\%  &14.09\%  &4.72\% &451~s  &64~s  &190~s  &30~s  &8~s \\ 
				Case 3 &30.32\%  &432.3\%  &51.45\%  &375.76\%  &6.94\% &433~s  &53~s  &177~s  &24~s  &8~s \\ 
				Case 4 &14.98\%  &53.23\%  &23.09\%  &18.83\%  &11.71\% &403~s  &95~s  &154~s  &50~s  &8~s \\ 
				\bottomrule
			\end{tabular}
		}
	\end{table*}
	
	The quantitative results in Table~\ref{tab:comparePDNN} further support the above observations. In Case 1 and Case 2, PDNN-Full achieves the lowest relative errors of 2.23\% and 3.84\%, respectively, but its runtime is 451 s in both cases. CRPDNN gives slightly higher errors of 4.16\% and 4.72\%, while reducing the runtime to 8 s. This indicates that CRPDNN achieves comparable accuracy with substantially lower computational time. The advantage of CRPDNN becomes more evident in the difficult cases. In Case 3, PDNN-Subregion and IPDNN fail due to unreliable subregion selection, resulting in extremely large relative errors of 432.3\% and 375.76\%, respectively. IPDNN-InitFull also suffers from a high error of 51.45\%, while CRPDNN maintains a much lower error of 6.94\%. In Case 4, CRPDNN achieves the lowest error of 11.71\%, whereas PDNN-Full, PDNN-Subregion, IPDNN-InitFull, and IPDNN yield errors of 14.98\%, 53.23\%, 23.09\%, and 18.83\%, respectively.
	
	These results indicate that subregion-based acceleration can reduce runtime, but its performance is sensitive to the reliability of the selected region. PDNN-Full avoids this issue by optimizing over the whole DOI, but the runtime is much higher. By avoiding both fast-algorithm initialization and explicit subregion selection, while using adaptive loss-weight adjustment, CRPDNN provides more stable reconstructions across different cases with competitive computational efficiency.
	
	\begin{figure}[!t]
		\centering
		\includegraphics[width = 0.5\linewidth]{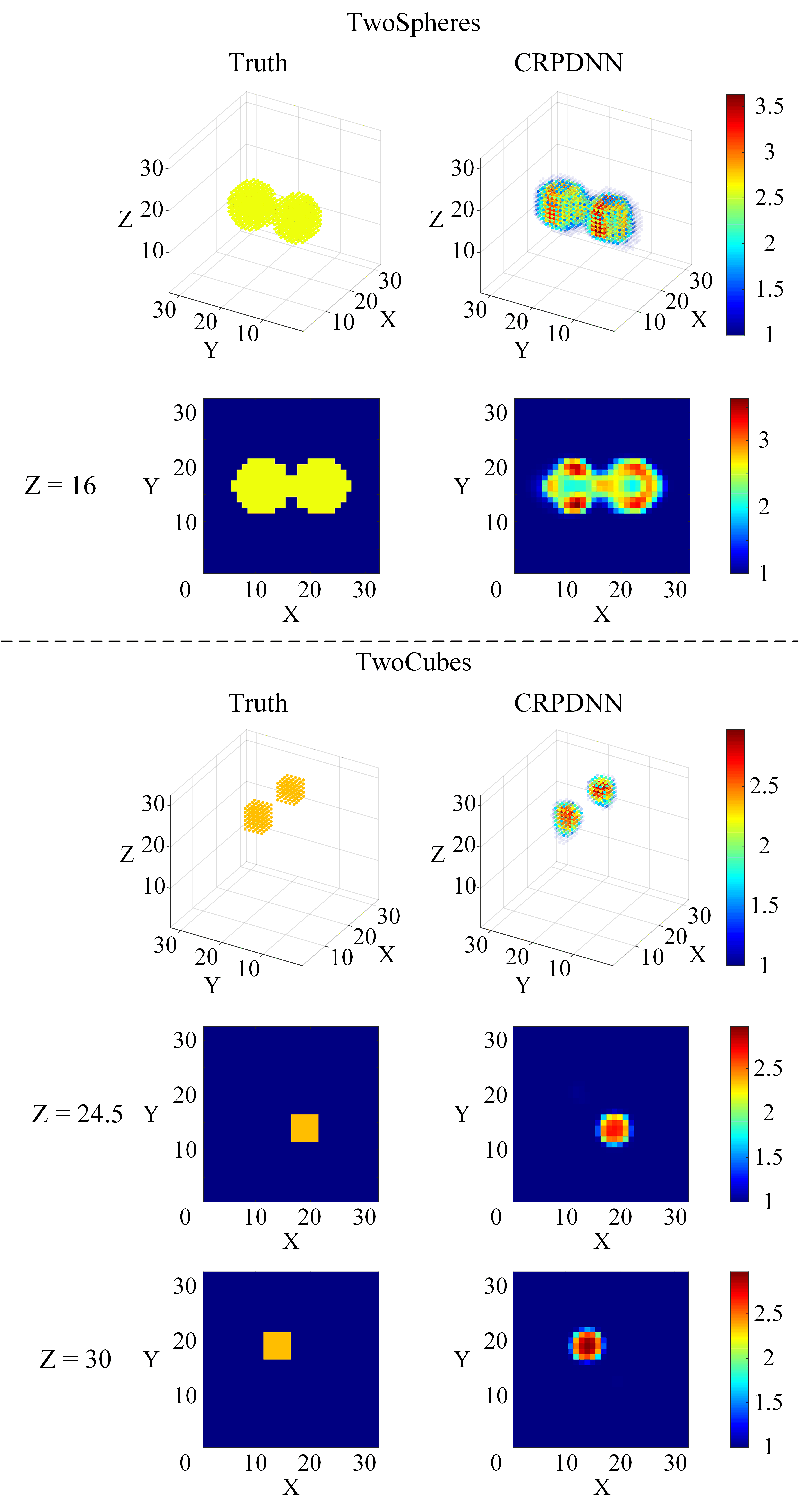}
		\caption{Volumetric and selected axial-slice reconstruction results of the 3-D Fresnel experimental targets ``TwoSpheres'' and ``TwoCubes'' using the proposed CRPDNN.}
		\label{fig:3DFresnel}
	\end{figure}
	
	\section{Experimental Validation}
	\label{sec:exp}
	To further evaluate the proposed solver on measured data, 3-D experiments from the Institut Fresnel free-space scattering database are considered \cite{Geffrin2005Fresnel, Geffrin2009Fresnel}. The co-polarized PP data are used, corresponding to the $\phi\phi$ polarization. The sources and receivers are located around the target at a radius of 1.796 m. The source positions cover azimuthal angles from $20^\circ$ to $340^\circ$ with a step of $40^\circ$ and polar angles from $30^\circ$ to $150^\circ$ with a step of $15^\circ$. Since parabolic antennas are used as transmitters, the incident field can be approximated as a plane wave within the target region. The receivers are located in the azimuthal plane, with azimuthal angles sampled from $0^\circ$ to $350^\circ$ with a step of $10^\circ$, excluding positions too close to the source direction.
	
	The ``TwoSpheres'' target consists of two dielectric spheres with a diameter of 50 mm and relative permittivity $\epsilon_r=2.6$. The ``TwoCubes'' target consists of two dielectric cubes with side length 25 mm and relative permittivity approximately $\epsilon_r=2.3$.
	
	Figure~\ref{fig:3DFresnel} presents the reconstruction results for the 3-D Fresnel experimental targets ``TwoSpheres'' and ``TwoCubes''. For ``TwoSpheres'', CRPDNN accurately recovers the locations and overall shape of the two dielectric spheres. Both the voxel visualization and the central slice show that the two adjacent spherical components are clearly identified, with only slight boundary smoothing and minor local fluctuations in the reconstructed permittivity. The relative reconstruction error is 2.54\%.
	
	For ``TwoCubes'', the proposed solver accurately reconstructs the two separated dielectric cubes at different axial positions. The slice views at different $z$ locations confirm that the reconstructed high-permittivity regions appear at the correct positions, and the background artifacts are well suppressed. Although the sharp cube edges are slightly smoothed, the overall geometry and permittivity contrast are well preserved. The relative reconstruction error is 0.59\%.
	
	These experimental results demonstrate that CRPDNN remains effective for 3-D measured scattering data and can achieve accurate volumetric reconstructions under realistic measurement conditions.
	
	\section{Conclusion}
	\label{sec:Conclu}
	
	This paper proposed a coordinate-residual physics-driven neural network (CRPDNN) for 3-D electromagnetic inverse scattering. The proposed solver directly represents the unknown volumetric contrast distribution in the voxel domain using a residual convolutional network with normalized spatial coordinates as part of the input. The network parameters are optimized by enforcing consistency between the predicted and measured scattered fields. Different from subregion-accelerated PDNN frameworks, CRPDNN does not require a preliminary reconstruction. In addition, data fidelity and regularization terms are balanced through adaptive loss-weight adjustment, reducing the dependence on fixed empirical hyperparameters.
	
	The proposed CRPDNN was validated on 3-D synthetic and Fresnel experimental ISPs, including digit-like scatterers, complex-valued targets, and noise-contaminated measurements. The results show that CRPDNN achieves more accurate reconstructions with fewer background artifacts, better structural consistency, and shorter runtime than CSI and $L_{2/3}$-FBE-WCIE, while the Fresnel results confirm its applicability to measured 3-D scattering data. A supplementary 2-D comparison with existing PDNN-based frameworks further shows that CRPDNN avoids the failure risk caused by inaccurate subregion selection and achieves a significantly shorter average runtime.
	
	Overall, the proposed solver provides a stable
	physics-driven reconstruction framework for 3-D electromagnetic inverse scattering. Future work will focus on extending the solver to more general measurement configurations and further reducing the computational cost for large-scale 3-D imaging problems.

	\bibliographystyle{unsrt}
	\bibliography{ref}
\end{document}